\documentclass{jpconf}
\usepackage{epsfig}
\usepackage[sort,numbers]{natbib}
\usepackage{amsmath}

\usepackage{graphicx}
\usepackage{subfigure}

\usepackage{hyperref}
\usepackage{float}

\begin{document}

\title{Radial Acceleration and Tully-Fisher Relations in Conformal Gravity}

\author{James~G.~O'Brien$^1$, Thomas~L.~Chiarelli$^2$, Philip~D.~Mannheim$^3$, Mark~A.~Falcone$^4$,  Muhannad~H.~AlQurashi$^4$, and Jordan Carter$^4$}

\address{$^1$Department of Mathematics, Physics and Computer Science, Springfield College, Springfield, MA
01109, USA}

\address{$^2$Department of Electromechanical Engineering, Wentworth Institute of Technology, Boston, MA
02115, USA}

\address{$^3$Department of Physics, University of Connecticut, Storrs, CT 06268, USA}

\address{$^4$Department of Mechanical Engineering, Wentworth Institute of Technology, Boston, MA
02115, USA}

\ead{jobrien7@springfieldcollege.edu,chiarellit@wit.edu,philip.mannheim@uconn.edu}
\date{December 5, 2018}

\begin{abstract}
In  2016 McGaugh, Lelli and Schombert established a universal Radial Acceleration Relation for centripetal accelerations in spiral galaxies.
Their work showed a strong correlation between observed centripetal accelerations and those predicted by luminous Newtonian matter alone. Through the use of the fitting function that they introduced, mass discrepancies in spiral galaxies can be constrained in a uniform manner that is completely determined by the baryons in the galaxies. Here we present a new empirical plot of the observed centripetal accelerations and the luminous Newtonian expectations, which more than doubles the number of observed data points considered by McGaugh et al. while retaining the Radial Acceleration Relation.  If this relation is not to be due to dark matter, it would then have to be due to an alternate gravitational theory that departs from Newtonian gravity in  some way. 
In this paper we show how the candidate alternate conformal gravity theory can provide a natural description of the Radial Acceleration Relation, without any need for dark matter or its free halo parameters. We discuss how the empirical Tully-Fisher relation follows as a consequence of conformal gravity.
\end{abstract}

\section{Introduction}
The rotation curves of spiral galaxies have been intensively studied for many years and serve as a key case study for the missing mass problem.  Following the pioneering work of Freeman \cite{freeman}, of Roberts and Whitehurst \cite{roberts}, and of Rubin, Ford and Thonnard \cite{rubin1980}, there has been extensive study of the rotation curves of spiral galaxies, in particular in the region far beyond the optical disk.  The cold dark matter (CDM) formalism, first developed for spiral galaxies by Navarro, Frenk and White (NFW) \cite{nfw}, explains mass discrepancies through the introduction of dark matter, to thus remain consistent with Newtonian gravity.  Even with the ability of NFW to fit rotation curves, a lack of direct observational evidence for dark matter and the shrinking allowed parameter space still allowed for it invites alternate explanations of the missing mass problem.  Some typical attempts are Modified Newtonian Dynamics (MOND) \cite{milgrom}, Modified Gravity (MOG) \cite{moffat1}, and Conformal Gravity (CG) \cite{mannheimfull}.  Typical of these alternate theories is the ability not just to fit galactic rotation curves without dark matter, but to do so without the two free parameters per galactic halo that are needed in CDM-type theories.  For the 236 galaxies considered here, dark matter theories currently require 472 free halo parameters, none of which are needed in the CG theory we study here.

Recently, McGaugh, Lelli and Schombert identified a universal relation between the observed centripetal accelerations in galaxies and the centripetal accelerations predicted by  luminous Newtonian matter alone \cite{mcgaughprl}.  
They reached this conclusion not by looking at individual rotation curves, but by plotting all of the centripetal accelerations on a single plot.  In \cite{mcgaughprl} it was found that the observed centripetal accelerations {$g_{OBS}=v^2_{OBS}/r$} are correlated with the  baryonic luminous Newtonian $g_{NEW}=v^2_{NEW}/r$ by a fitting function
\begin{equation}
g_{OBS}=\frac{g_{NEW}}{[1-\exp(-(g_{NEW}/g^{\dagger})^{1/2})]},
\label{E1}
\end{equation}
where {$g^{\dagger}$} is a fitting parameter with a fitted value of {$g^{\dagger}=1.2\times10^{-10}ms^{-2}$}.  This correlation has been called the Radial Acceleration Relation (RAR).

By being able to include all data points in a single plot, the empirical study of McGaugh et al. is more extensive and comprehensive than previous empirical studies such as that of Persic, Salucci and Stel \cite{Persic}.  Implicit in the study of McGaugh et al. is the suggestion that there is a universal nature to the pattern of accelerations observed in spiral galaxies.  As such, McGaugh et al. provide a universal formula which can be used to test and/or constrain any theory of galactic rotation curves.  Since the RAR is an analysis of every data point in each rotation curve, it is more general than the well known Tully-Fisher (TF) relation, which is found to hold in many spiral galaxies \cite{Tully1997a} as
\begin{equation}
  v_{OBS}^4\propto M,
 \label{tf}
 \end{equation}
where $M$ is the total visible mass in each galaxy.
 As such, both the RAR and TF relation challenge any theory of galactic rotation curves.  As noted by McGaugh et al. themselves, eq. (\ref{E1}) either constrains and challenges dark matter theory or implies alternate physics that goes beyond the standard Newton-Einstein theory.  In this paper we discuss a particular alternate theory, namely conformal gravity, and show how it describes the RAR and the TF relation in a universal fashion.

\section{Conformal Gravity Rotation Curve Fitting}

By being a pure metric theory of gravity, conformal gravity, just like Einstein gravity, identifies gravity as spacetime curvature.  However, unlike Einstein gravity, conformal gravity  imposes an additional conformal invariance symmetry.  In consequence, conformal gravity is derived from an action based on the square of the Weyl tensor as
\begin{equation}
I_{\rm W}=-\alpha_g\int d^4x (-g)^{1/2}C_{\lambda\mu\nu\kappa}
C^{\lambda\mu\nu\kappa},
\label{182}
\end{equation}
where 
\begin{equation}
C_{\lambda\mu\nu\kappa}= R_{\lambda\mu\nu\kappa}
-\frac{1}{2}\left(g_{\lambda\nu}R_{\mu\kappa}-
g_{\lambda\kappa}R_{\mu\nu}-
g_{\mu\nu}R_{\lambda\kappa}+
g_{\mu\kappa}R_{\lambda\nu}\right)
+\frac{1}{6}R^{\alpha}_{\phantom{\alpha}\alpha}\left(
g_{\lambda\nu}g_{\mu\kappa}-
g_{\lambda\kappa}g_{\mu\nu}\right)
\label{180}
\end{equation}
is the conformal Weyl tensor with dimensionless coupling constant $\alpha_g$. 
The resulting field equations, {$4\alpha_g\left(
2\nabla_{\kappa}\nabla_{\lambda}C^{\mu\lambda\nu\kappa}-
C^{\mu\lambda\nu\kappa}R_{\lambda\kappa}\right)=T^{\mu\nu}$} were solved by Mannheim and Kazanas in the region exterior to a static, spherically symmetric source \cite{Mannheim1989}, where it was shown that a point stellar mass produces a potential $V^*(r)=-\beta^* c^2 /r+\gamma^* c^2 r/2$.  To go from the single star  to the prediction for rotational velocities of entire spiral galaxies,
it was shown \cite{fitting} that the resulting galactic velocity expectation is given by
\begin{equation}
v_{CG}(R) = \left[v_{NEW}^2(R)+\frac{M}{M_\odot}\frac{\gamma^*c^2R^2}{2R_0}I_1\left(\frac{R}{2R_0}\right)K_1\left(\frac{R}{2R_0}\right)+\frac{\gamma_0c^2R}{2}-\kappa c^2R^2\right]^{1/2},
\label{total}
\end{equation}
where M is the mass of the galaxy in solar mass units ($M_\odot$), $R_0$ is the galactic disk scale length, and $v_{NEW}(R)$ is the standard Freeman formula for a Newtonian disk: 
\begin{equation}
v_{NEW}(R) = \left[\frac{M}{M_\odot}\frac{\beta^*c^2R^2}{2R^3_0}\left[I_0\left(\frac{R}{2R_0}\right)K_0\left(\frac{R}{2R_0}\right)-I_1\left(\frac{R}{2R_0}\right)K_1\left(\frac{R}{2R_0}\right)\right]\right]^{1/2}.
\label{gr}
\end{equation}
Fits to 236 galaxies have been achieved through eq. (\ref{total}) with the following fixed parameters: $\beta^*=1.48\times 10^5cm$, ${\gamma^*=5.42\times10^{-41} {cm}^{-1}}$, $\gamma_0=3.06\times10^{-30} {cm}^{-1}$ and
$\kappa=9.54\times10^{-54}~{cm}^{-2}$ \cite{fitting}.  In actual applications of eq. (\ref{total}), one typically includes gas contributions as well as bulge contributions (when applicable). For  details on the specifics, see \cite{fitting}.

In recent years CG  has enjoyed  success in fitting rotational velocities of spiral galaxies without dark matter.  New physics is encapsulated in the uniquely CG terms in eq. (\ref{total}), terms that depend on just three new parameters $\gamma^*$, $\gamma_0$ and $\kappa$ beyond the standard Newtonian $\beta^*$, with no fine tuning of input parameters being required.  To date, all observed spiral galaxy rotation curves have been explained by conformal gravity in a universal manner despite  varying morphologies, luminosities and masses of galaxies studied.  Although McGaugh et al. provide a new analysis of rotation curves, it should be noted that the issue of universality in centripetal accelerations of spiral galaxies has been discussed previously in conformal gravity.  Specifically, the values of $(v_{OBS}^2/R)_{last}$ for the last data points in each and every studied galaxy were found to be of the order of $3\times 10^{-11}ms^{-2}$, with $(v_{OBS}^2/c^2R)_{last}$ being of order $3\times 10^{-30}cm^{-1}$ (see for example, \cite{fitting,impact} and the last column in Table \ref{t1}).  In the next section we will discuss the procedure for reconstructing the plots provided by McGaugh et al. while also extending their data sample substantially by using the 236 galaxies that CG has successfully fit.

\section{Reproducing the Study of McGaugh, Lelli and Schombert}

The analysis of McGaugh et al. uses the Spitzer Photometry and Accurate Rotation Curves (SPARC) database \cite{sparc}, which consists of 175 spiral galaxies, containing 3149 total rotation curve data points.  The points {$(R,v_{OBS})$} are the observed velocity {$v_{OBS}$} at a given distance from the galactic center {$R$}.  Using these points one can calculate the observed centripetal accelerations as {$g_{OBS}=v^2_{OBS}/R$}.  In order to calculate $g_{NEW}$, McGaugh et al. used 
fixed mass to light ratios of 0.5 for disks and 0.7 for bulges.  After filtering the data, the net number of galaxies used by McGaugh et al. was reduced to 153, with a net total of 2693 points \cite{onelaw}.  With $g_{NEW}$ being calculated from luminous Newtonian matter alone, and $g_{OBS}$ being based on the observed velocities, McGaugh et al. constructed a plot of $g_{OBS}$ versus $g_{NEW}$  for every data point in their sample.  In such a plot, any deviation form the line of unity, namely the line of $g_{OBS} = g_{NEW}$, would reflect a mass discrepancy.  The resulting 2693 point plot as given in \cite{mcgaughprl} has exactly the same structure as the 2693 point Fig. 1(a) plot that we provide here, as well as having the same structure as the Fig. 1(b) plot  that uses the entire 6377 data points considered here. In both Fig. 1(a) and Fig. 1(b) we have plotted eq. (\ref{E1}), to show that it provides a good characterization of the data.

 \begin{figure}[H]
  \subfigure[]{%
 \epsfig{file=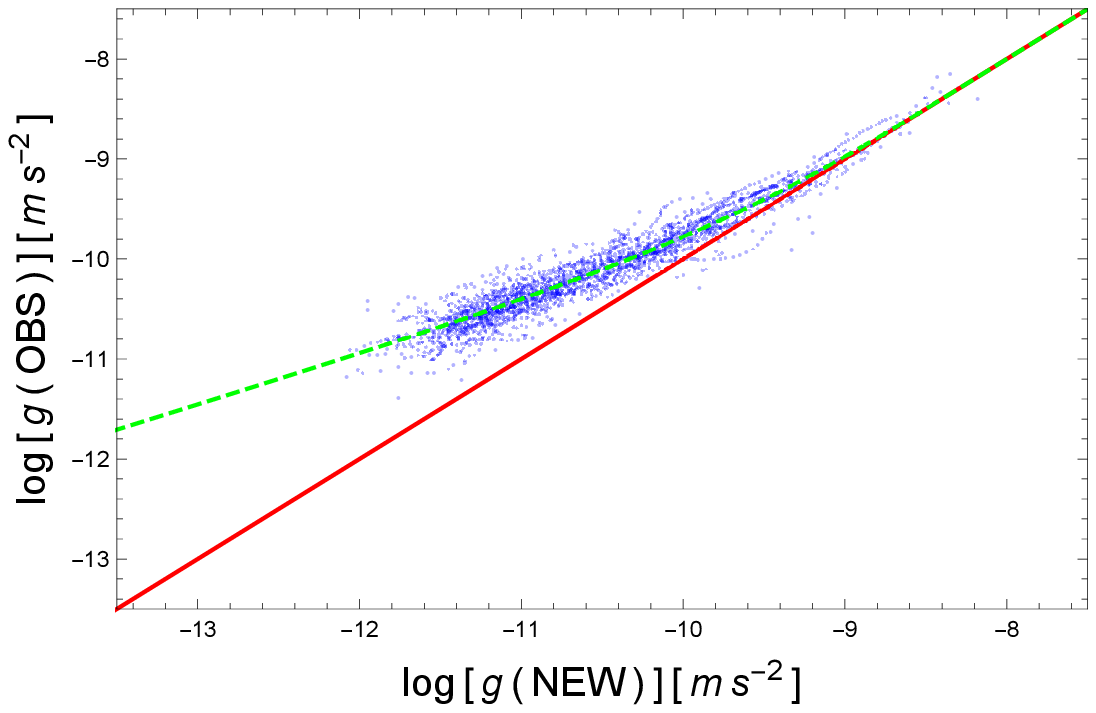,width=3.2in,height=2.05in}
  \label{complot}}
  \subfigure[]{%
 \epsfig{file=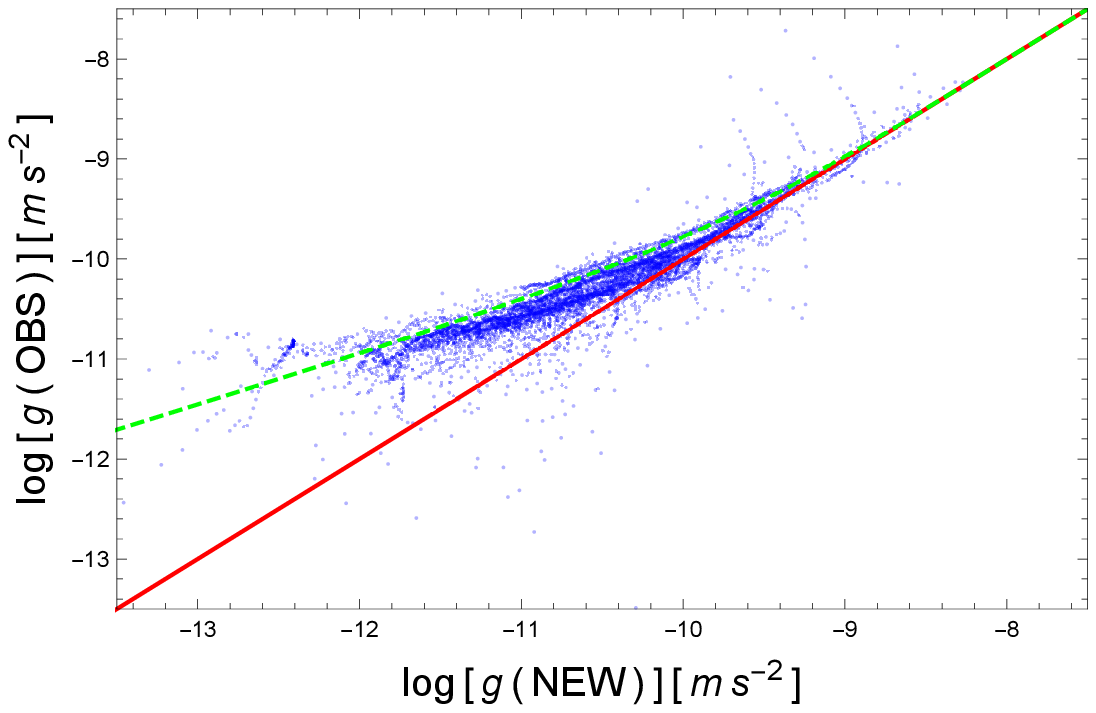,width=3.2in,height=2.05in}
  \label{ourplot}}
  \caption{ Fig. \ref{complot} shows the ($g_{OBS},g_{NEW}$) plot using the 2693 points studied by McGaugh et al.  Fig. \ref{ourplot} shows the ($g_{OBS},g_{NEW}$) plot using the full 6377 points.  The solid  line is the line of unity $g_{OBS}=g_{NEW}$, and the dashed line in both Fig. \ref{complot} and Fig. \ref{ourplot} is  eq. (\ref{E1}).}
   \label{dataplots}
\end{figure}

\section{Our Data Sample}
\begin{figure}[H]
\epsfig{file=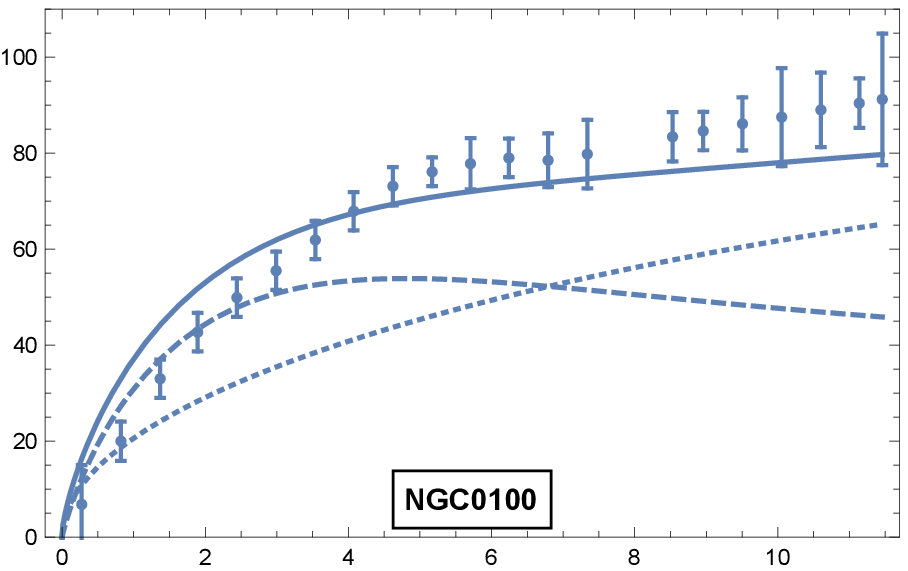,width=1.5825in,height=1.2in}~~~
\epsfig{file=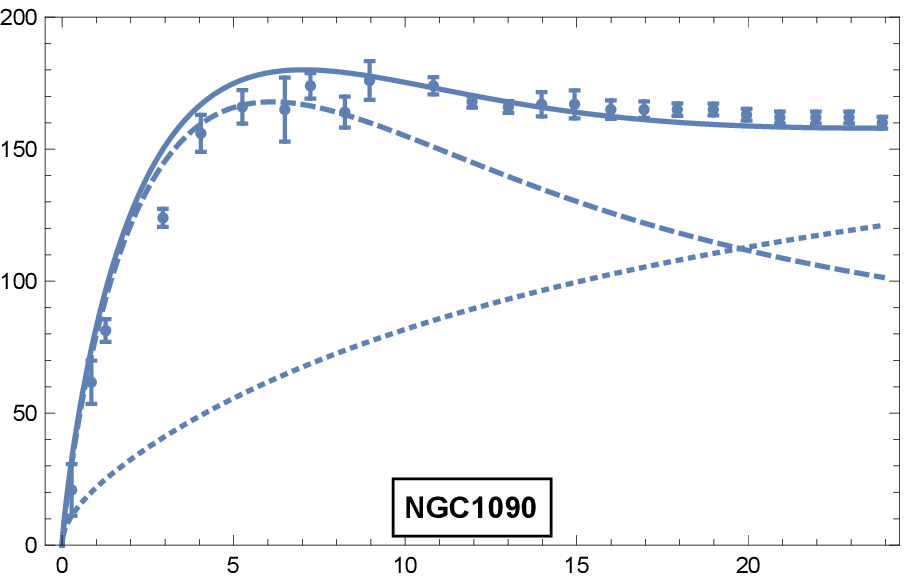,width=1.5825in,height=1.2in}~~~
\epsfig{file=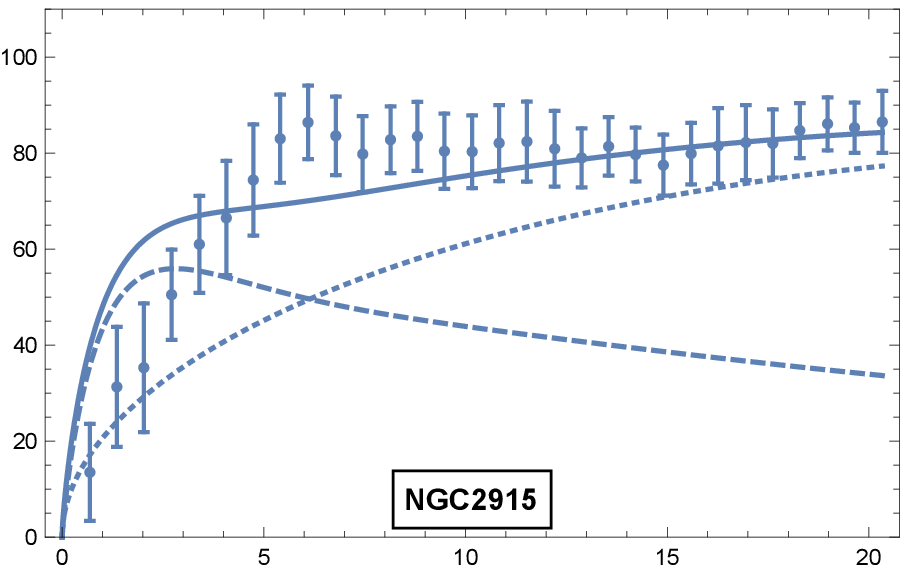,width=1.5825in,height=1.2in}~~~
\epsfig{file=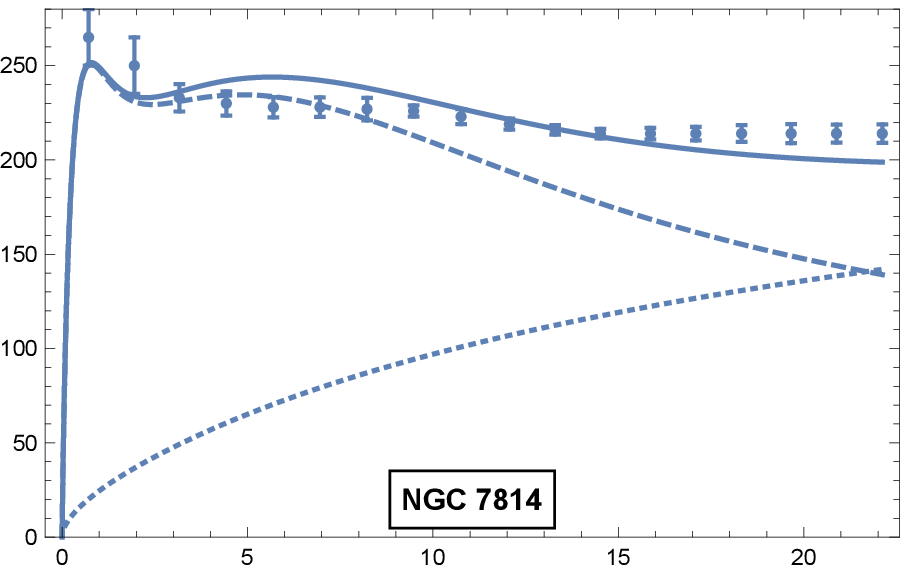,width=1.5825in,height=1.2in}\\
\smallskip
\epsfig{file=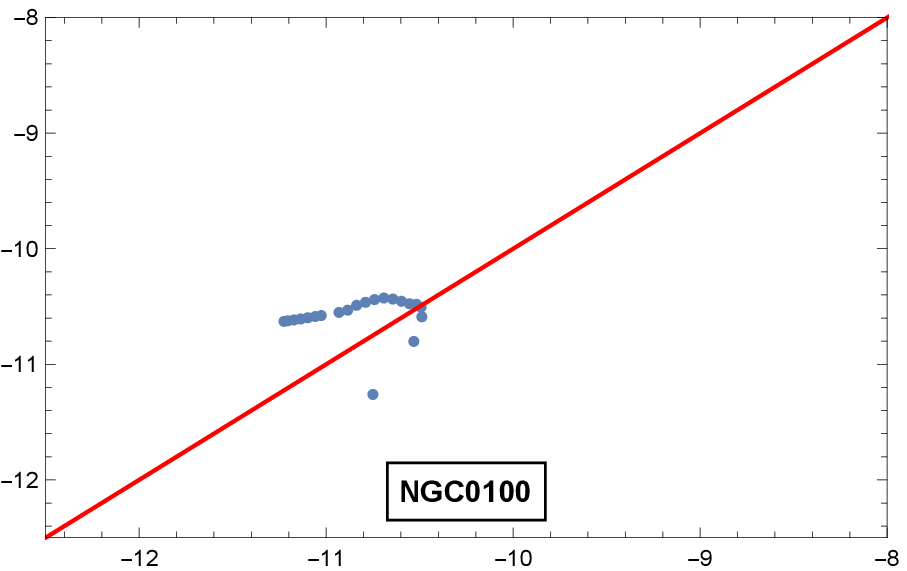,width=1.5825in,height=1.2in}~~~
\epsfig{file=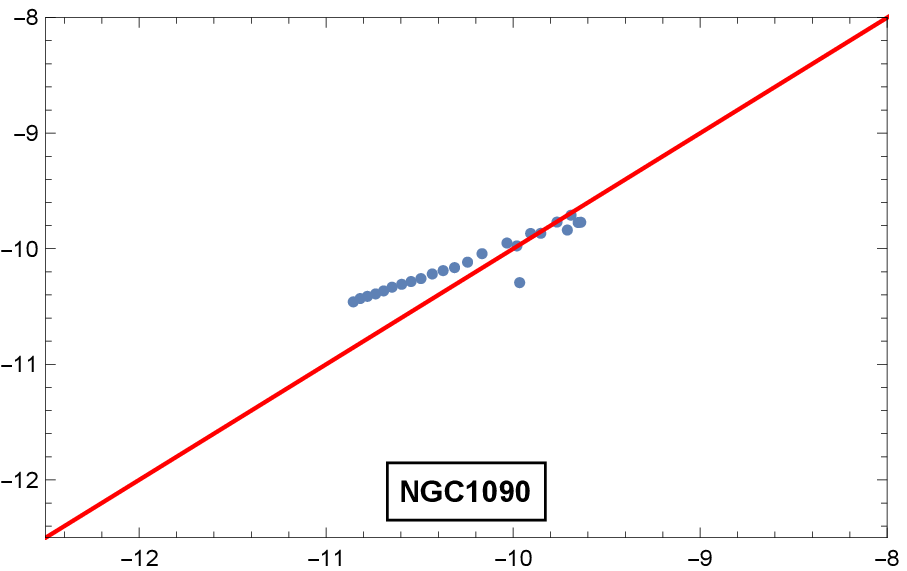,width=1.5825in,height=1.2in}~~~
\epsfig{file=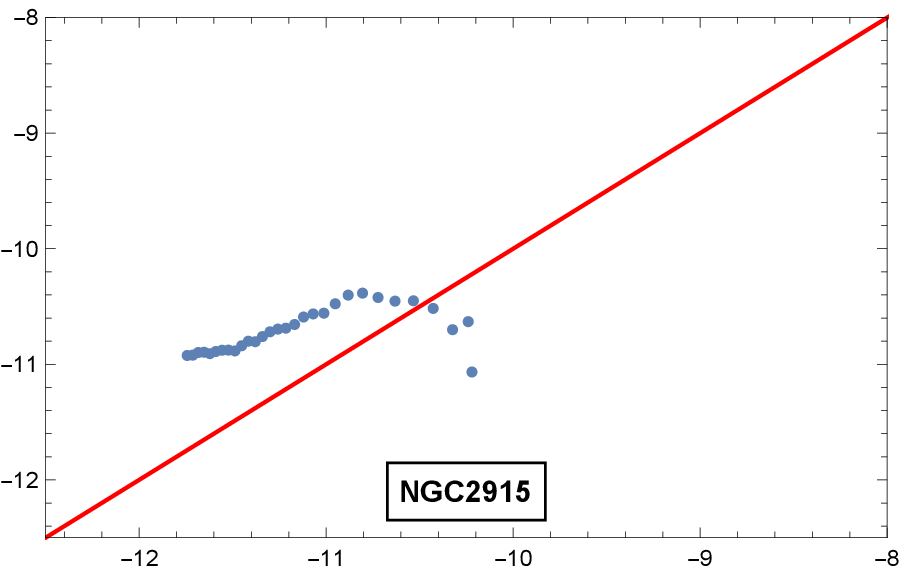,width=1.5825in,height=1.2in}~~~
\epsfig{file=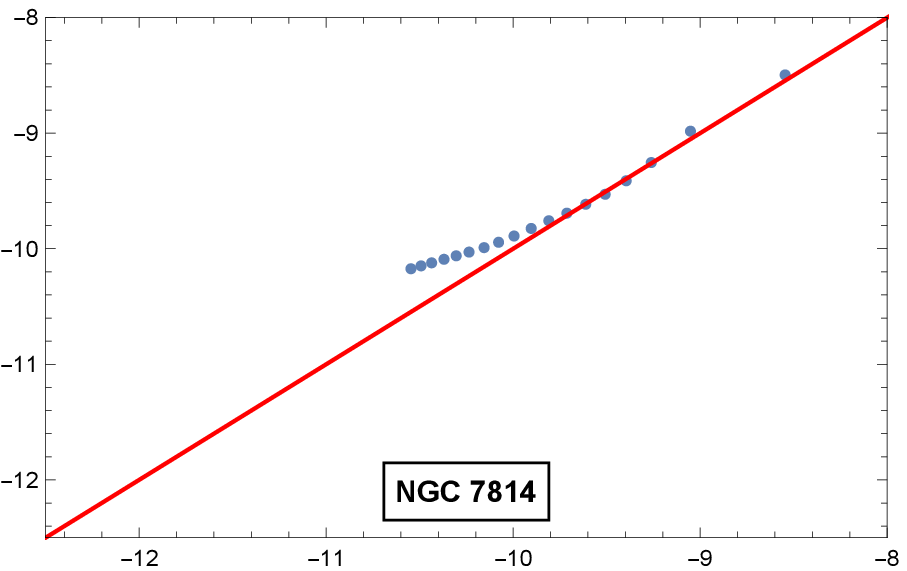,width=1.5825in,height=1.2in}\\
\smallskip
\epsfig{file=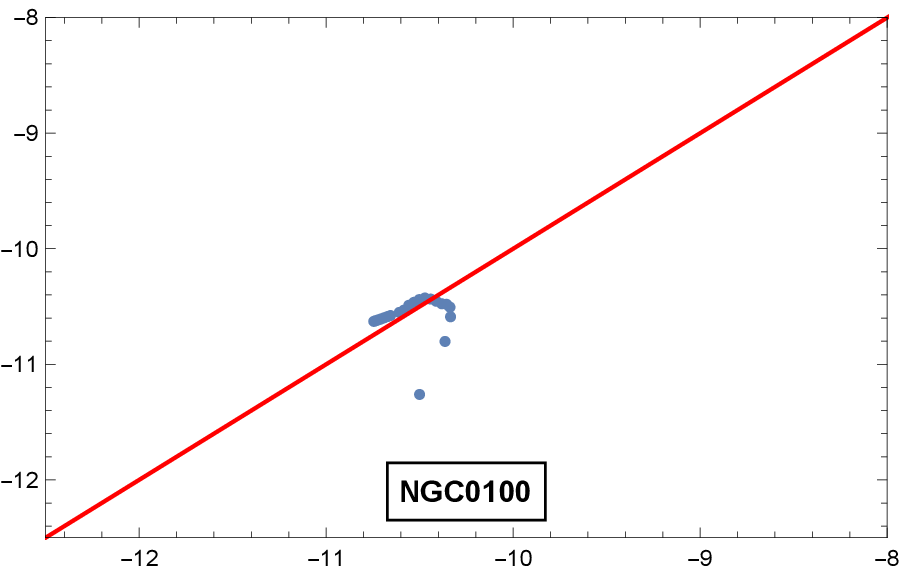,width=1.5825in,height=1.2in}~~~
\epsfig{file=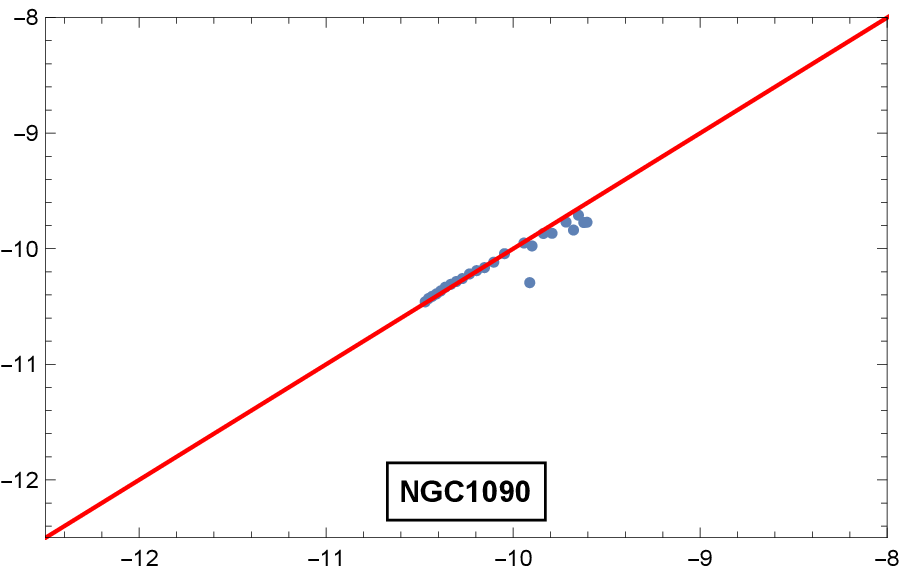,width=1.5825in,height=1.2in}~~~
\epsfig{file=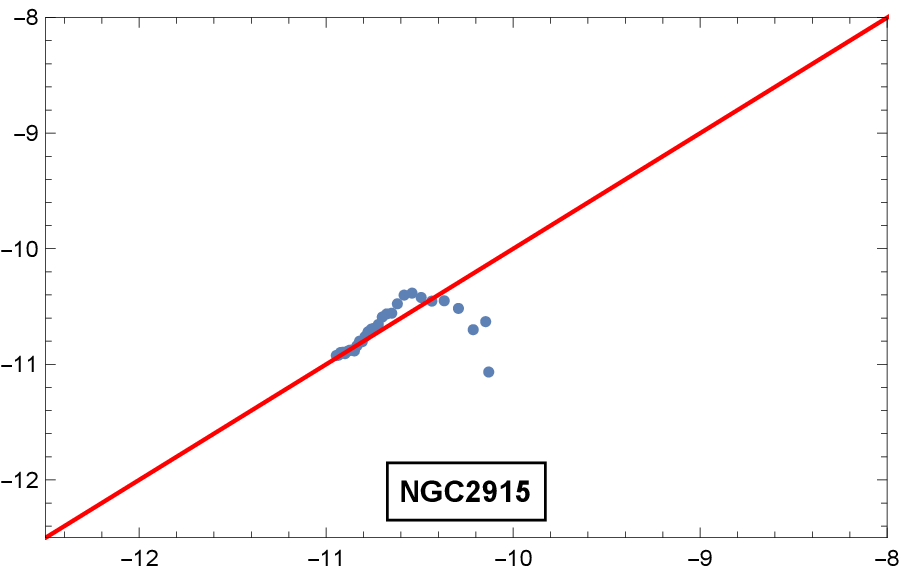,width=1.5825in,height=1.2in}~~~
\epsfig{file=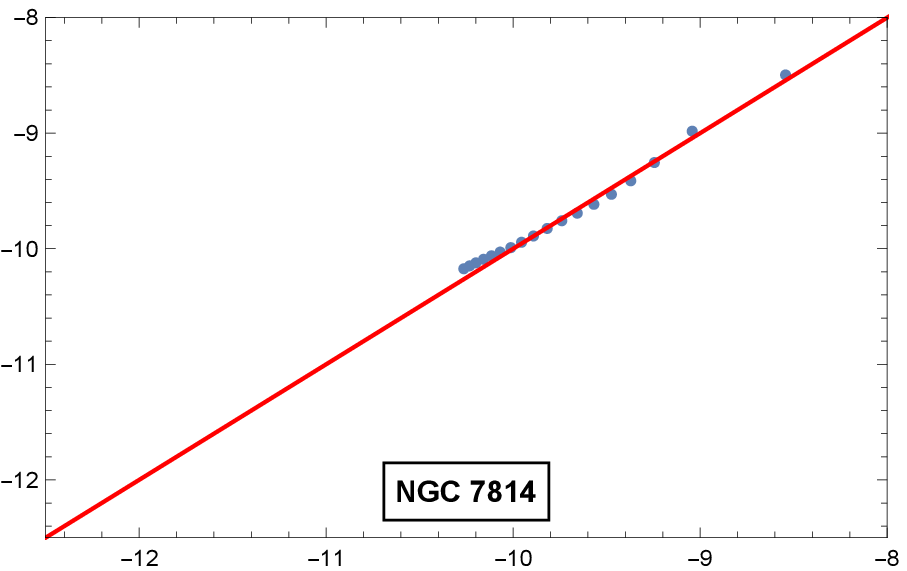,width=1.5825in,height=1.2in}\\
\medskip
\epsfig{file=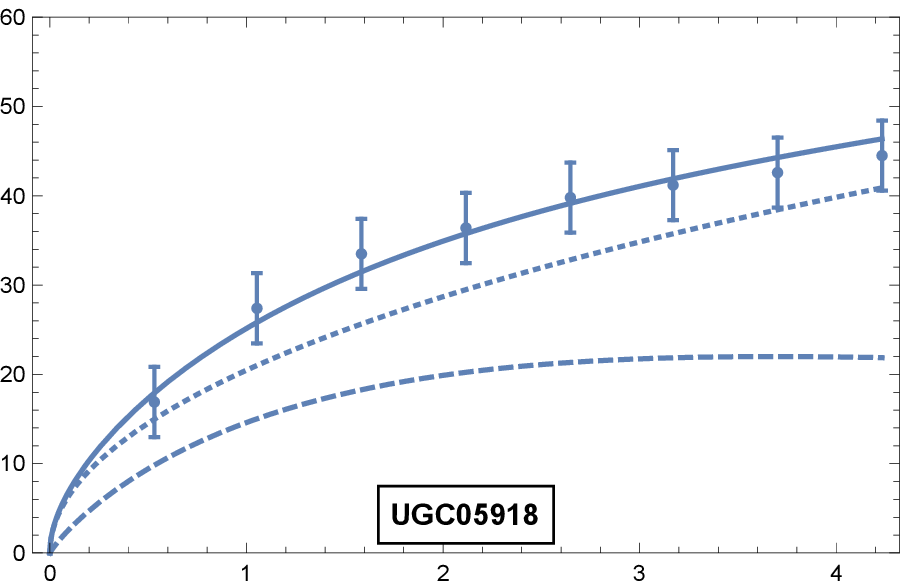,width=1.5825in,height=1.2in}~~~
\epsfig{file=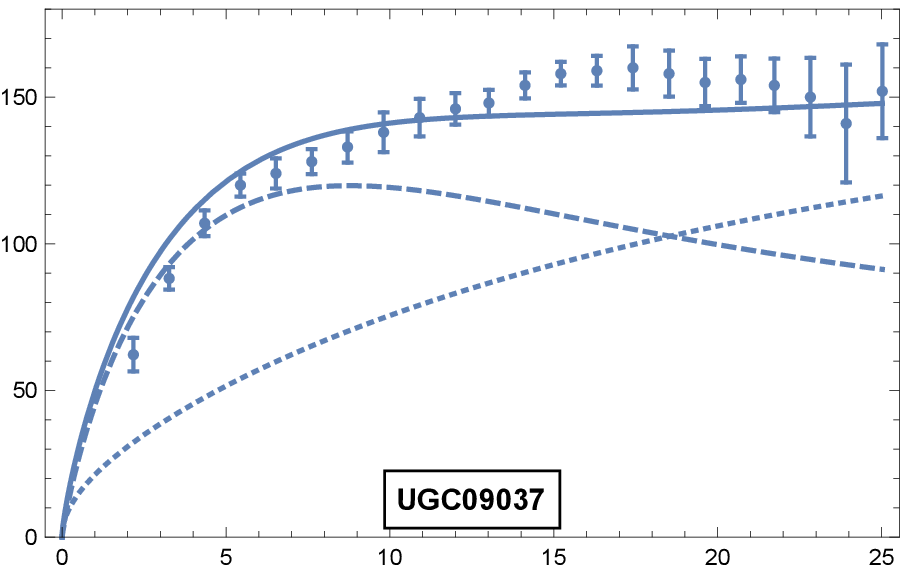,width=1.5825in,height=1.2in}~~~
\epsfig{file=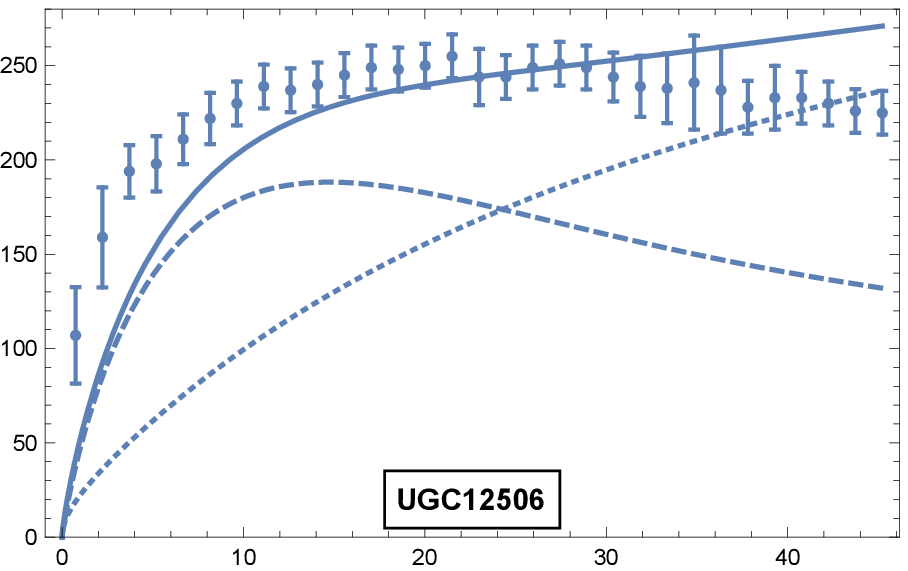,width=1.5825in,height=1.2in}~~~
\epsfig{file=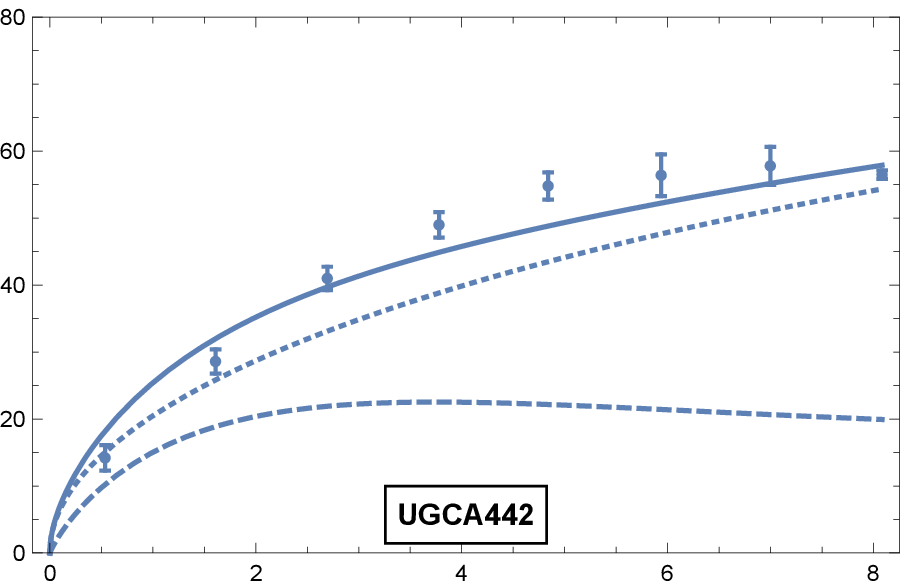,width=1.5825in,height=1.2in}\\
\smallskip
\epsfig{file=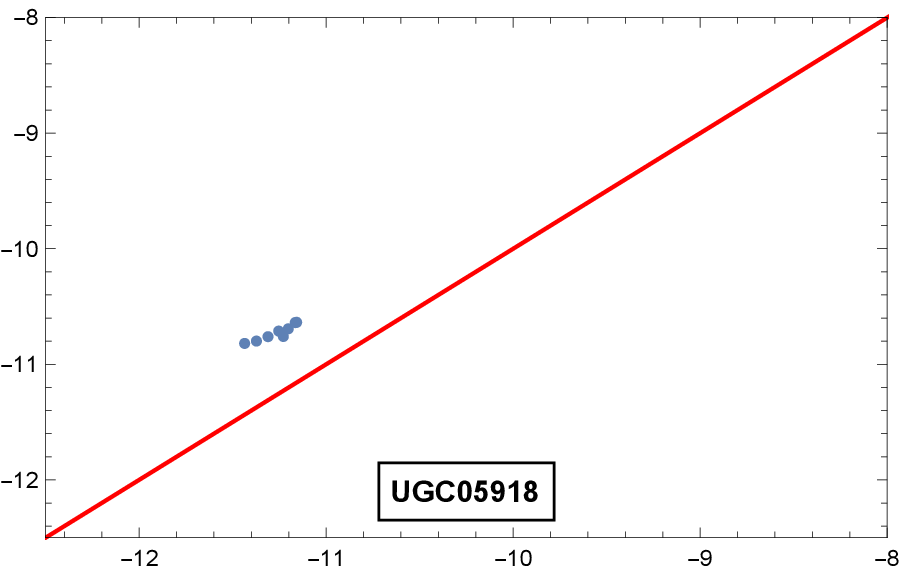,width=1.5825in,height=1.2in}~~~
\epsfig{file=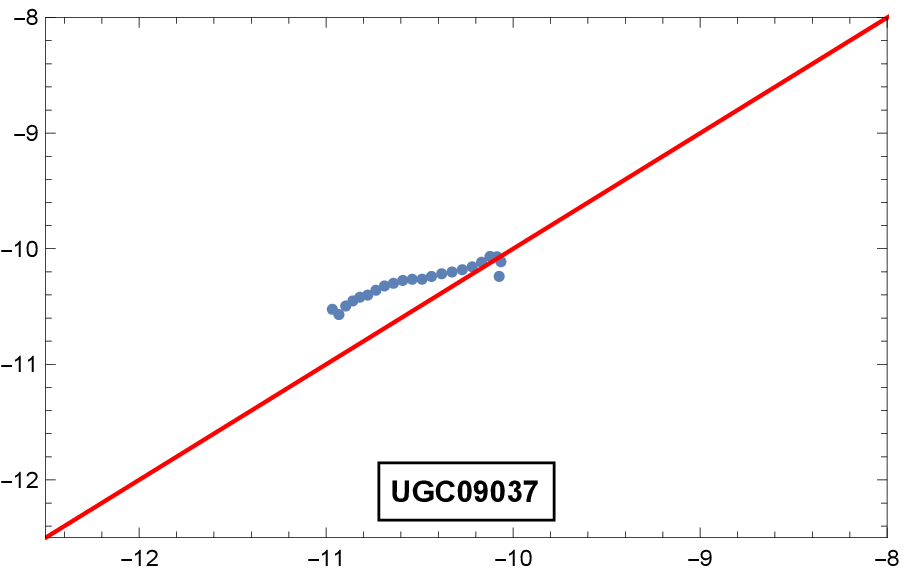,width=1.5825in,height=1.2in}~~~
\epsfig{file=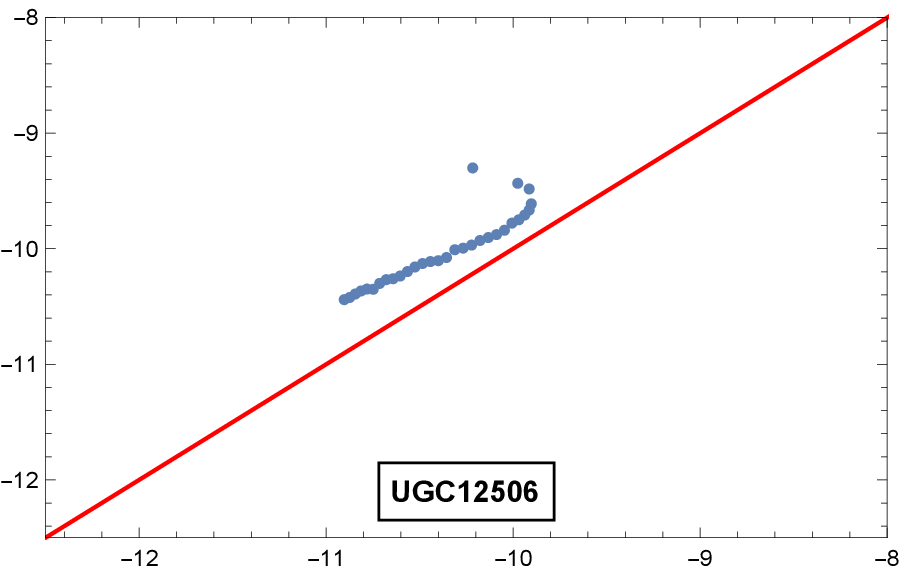,width=1.5825in,height=1.2in}~~~
\epsfig{file=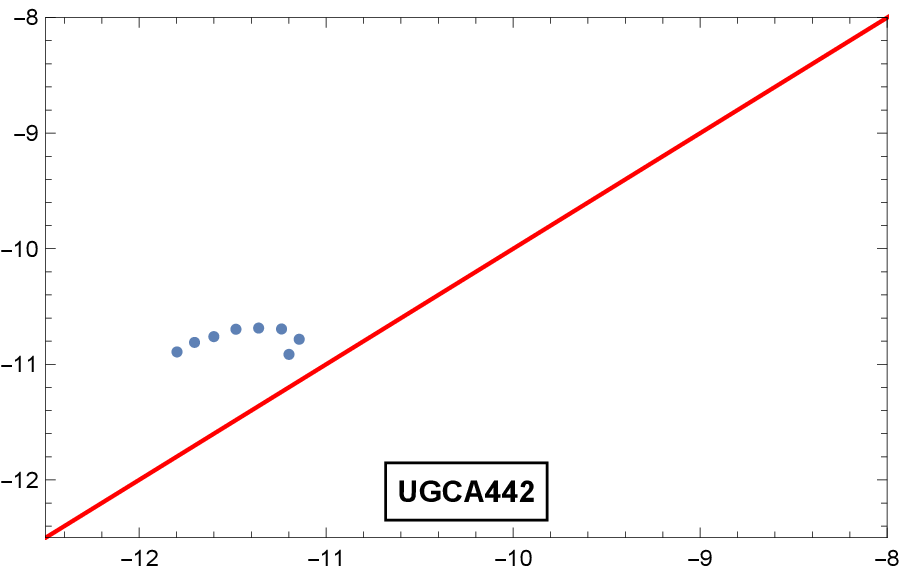,width=1.5825in,height=1.2in}\\
\smallskip
\epsfig{file=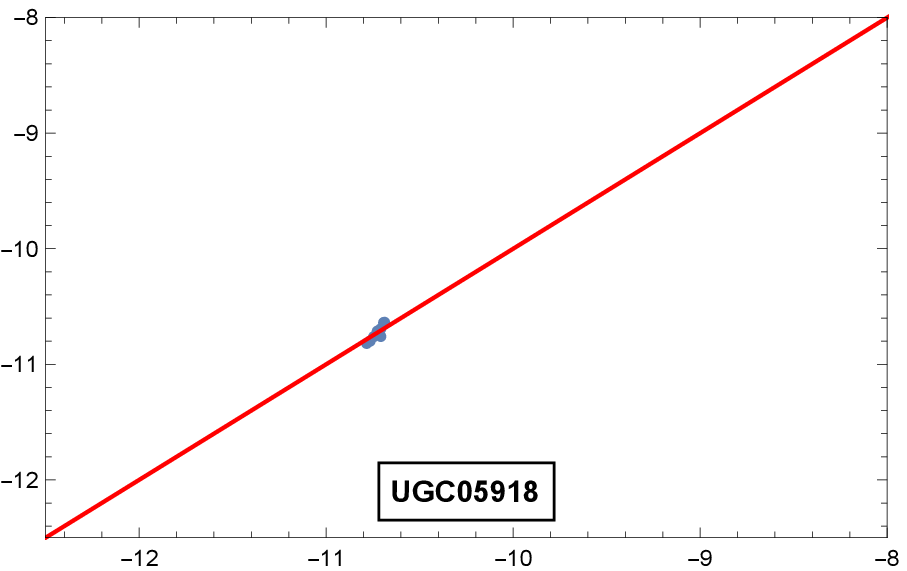,width=1.5825in,height=1.2in}~~~
\epsfig{file=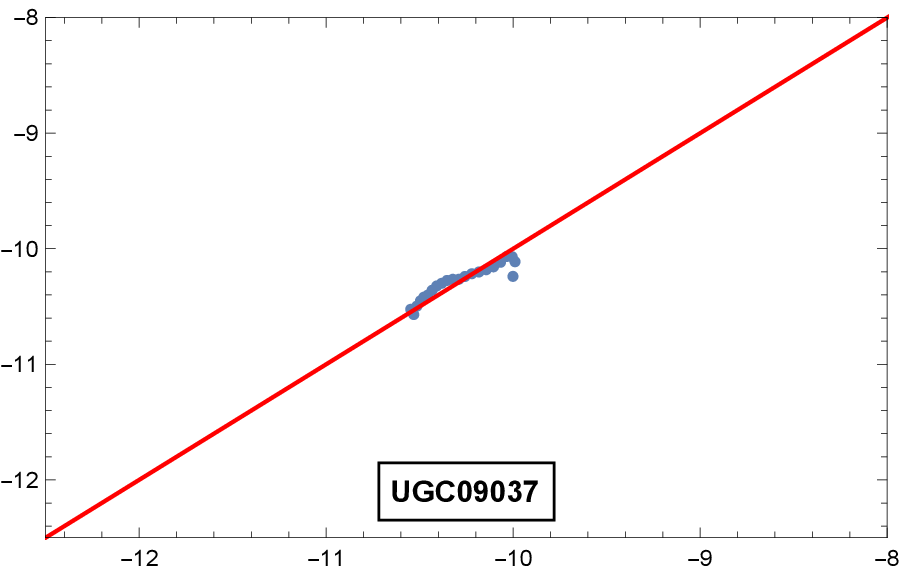,width=1.5825in,height=1.2in}~~~
\epsfig{file=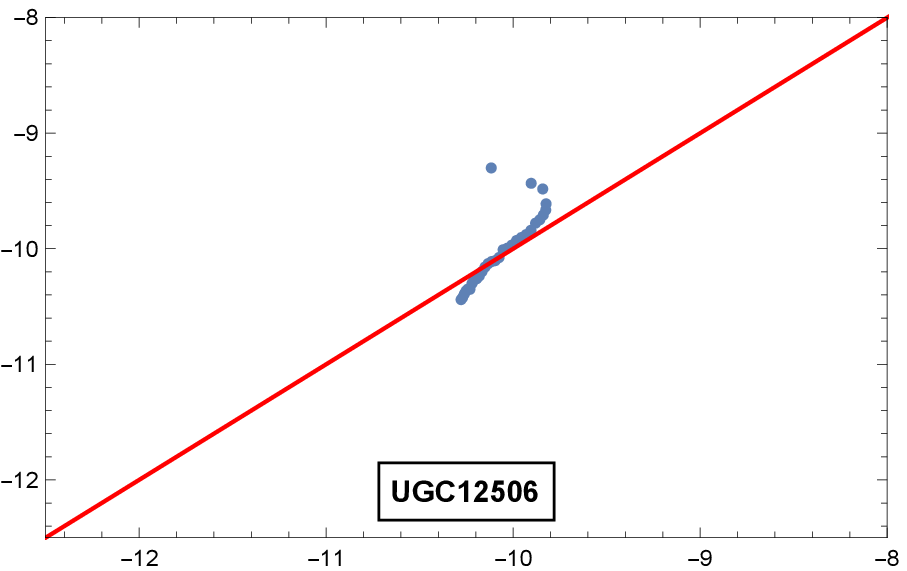,width=1.5825in,height=1.2in}~~~
\epsfig{file=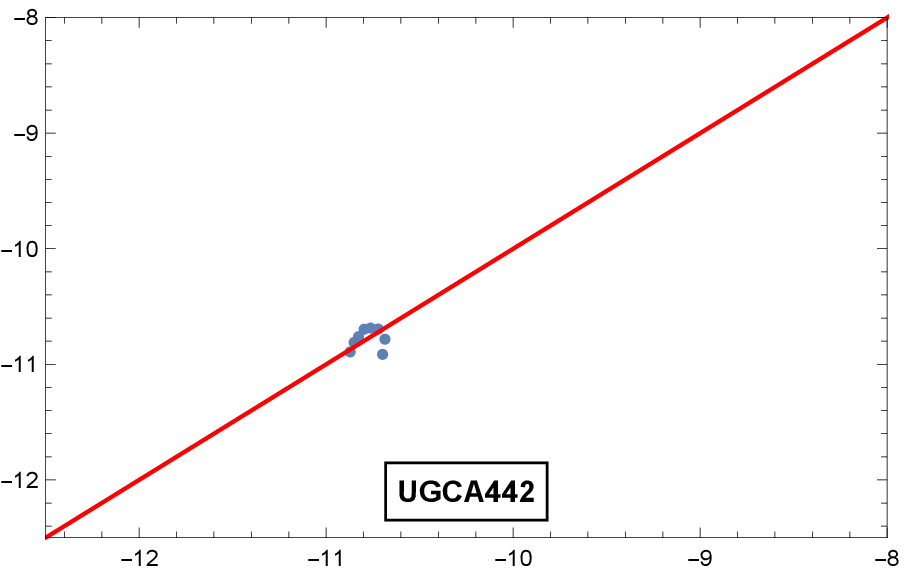,width=1.5825in,height=1.2in}\\
\smallskip

\caption{Fitting to the rotational velocities (in ${\rm km}~{\rm sec}^{-1}$) of the selected eight galaxy sample with their reported errors as plotted as a function of radial distance (in ${\rm kpc}$) is given in rows one and four. The dashed line shows the luminous Newtonian contribution, the dotted line is the contribution of the $\gamma^*$, $\gamma^0$ and $\kappa$ terms in eq. (\ref{total}), and the full curve is the conformal gravity fit.  Rows two and five show the respective contributions of the eight galaxies to Fig. \ref{ourplot}, and rows three and six show their respective contributions to Fig. \ref{opacity}.}
\label{rotcurves}
\end{figure}

Since the work of McGaugh et al. provides a universal description of galactic rotation curves, it is not specific to the data set that they chose, and not only should it not be, as we see it is reflected in the larger data set we study.  Our data sample includes all of the galaxies currently fit by CG in the past decade, as well as any galaxies that were included in \cite{onelaw} that were not yet fit by CG.  Data sets previously fit by CG include surveys such as  THINGS  \cite{Walter2008}, LITTLE THINGS  \cite{lthings},  the Ursa Major survey \cite{Verheijen1999}, and the WHISP Dwarfs survey \cite{Swaters2002a}. It should be noted that many of these galaxies in the aforementioned samples are galaxies already included in the SPARC database.  However, there are some galaxies that were included in the SPARC database that were not previously fit by CG. To this end, fits were performed on the remainder of the SPARC database, including the surveys of Spekkens et al. \cite{sg06}, Lelli et al. \cite{Lelli14}, Walsh \cite{Walsh97}, DeBlok et al. \cite{deBlok1996}, and Noordermeer et al. \cite{noord}, to thereby complete the CG fitting of the entire SPARC database.  A typical selection of these new fits is shown in Fig. \ref{rotcurves}, and the entire set of these as yet unpublished conformal gravity rotation curve fits will be presented in a future work.  The relevant parameters for producing the fits in Fig. \ref{rotcurves} are included in Table \ref{t1}.  
\begin{table}[H]
\caption{Properties of typical galaxies in the SPARC data set.}
\centering
\begin{tabular}{l c c c c c c c c c} 
\hline\hline
Galaxy &$ i$        & Dis.    & Lum.& $R_0$ & $R_{last}$  &$M_{HI}$  	    & $M_{d}$        & M/L & $(v^2 / c^2 R)_{last}$  	 \\
\phantom{0000000000}   & $(o)$  & (Mpc) & ($10^9L_\odot$) 	& (kpc) & (kpc)  & ($10^9$$M_\odot)$ & ($10^9$$M_\odot)$ & $(M_\odot L_\odot^{-1})$	 & $(10^{-30}cm^{-1})$   \\
\hline 

    NGC0100 & 89    & 16.08 & \phantom{00}4.59  & 1.98  & 11.46 & \phantom{0}3.95  & \phantom{00}2.99  & \phantom{0}0.65  & 2.61  \\
    NGC1090 & 64    & 29.40 & \phantom{0}45.48 & 2.80  & 23.91 & \phantom{0}7.76& \phantom{0}46.13 & \phantom{0}1.01  & 3.86   \\
    NGC2915 & 56    & \phantom{0}8.22  & \phantom{00}2.63  & 1.11  & 20.33 & \phantom{0}2.92  & \phantom{00}1.76  & \phantom{0}0.67  & 1.33   \\
    NGC7814 & 89    & 16.31 & \phantom{0}95.56 & 2.88  & 22.12 & \phantom{0}1.92  & \phantom{0}74.70 & \phantom{0}0.93  & 7.46   \\
    UGC05918 & 46    & \phantom{0}7.27  & \phantom{00}0.21  & 1.58  & \phantom{0}4.23  & \phantom{0}0.37  & \phantom{00}0.41  & \phantom{0}1.97  & 1.68 \\
    UGC09037 & 65    & 74.81 & \phantom{0}54.95 & 3.83  & 25.02 & 21.39 & \phantom{0}30.47 & \phantom{0}0.55  & 3.32   \\
    UGC12506 & 86    & 90.99 & 114.18 & 6.68  & 45.21 & 40.72 & 136.26 & \phantom{0}1.19  & 4.03   \\
    UGCA442 & 64    & \phantom{0}5.55  & \phantom{00}0.23  & 1.51  & \phantom{0}8.08  & \phantom{0}0.60  & \phantom{00}0.39  & \phantom{0}1.71  & 1.42   \\
\hline
\end{tabular}
\label{t1}
Table Columns: Galaxy name, reported inclination, NED distance, total B-Band luminosity, disk scale length, distance to last data point, HI gas mass, disk mass, mass to light ratio,  centripetal acceleration of the last data point.
\end{table}

To go from the total 236 rotation curves to the centripetal acceleration plots, we follow the same procedure outlined in \cite{onelaw}, albeit with some important differences.  The first major difference is that for the CG fitting we  did not fix mass to light ratios to a particular value as McGaugh et al. had done.  Instead, we used the fitting procedure outlined in \cite{fitting} itself, and used the fitted masses that were obtained in those works.  Table 1 shows that this procedure produces acceptable mass to light ratios of the order of the mass to light ratio in the local solar neighborhood. The fitted masses were then used to produce the values for $g_{NEW}$ exhibited in Fig. 1.  A second significant difference in constructing our plots versus those found in \cite{onelaw} is our use of the NASA Extragalactic Database (NED) average distance for each galaxy, as it provides a common baseline. For both  individual rotation curve points and fitted mass to light ratios there is great sensitivity  to the adopted distance as most parameters scale with distance.  The last significant difference is our inclusion of the recent ultra high resolution rotation curves of the THINGS and LITTLE THINGS galaxies.  Some of the galaxies in these two surveys were included in \cite{onelaw} but older data sets were used instead of the work of \cite{Walter2008} and \cite{lthings}. Since the THINGS and LITTLE THINGS surveys represent some of the highest quality rotation curve data, we choose to include them in our analysis. All other considerations in individual fits such as inclusion of gas and bulge when applicable remain consistent with our previous studies.  In total our full data sample contains 236 galaxies with 6377 data points, thus significantly increasing the scope of the study of McGaugh et al.

\section{The Conformal Gravity Fits}
\label{disc}
\begin{figure}[H]
 \centering
 \epsfig{file=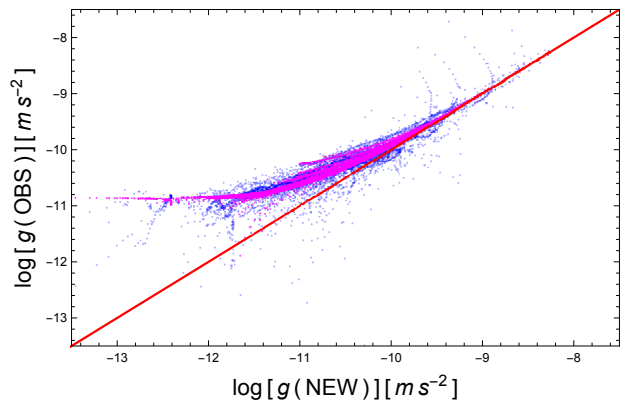,width=3.2in,height=2.05in}
  \caption{The ($g_{OBS},g_{NEW}$) plot with the  $g_{CG}$ prediction overlaid.  The solid line is the line of unity $g_{OBS}=g_{NEW}$. }
   \label{cplot}
\end{figure}

\begin{figure}[H]
	\subfigure[]{%
		\epsfig{file=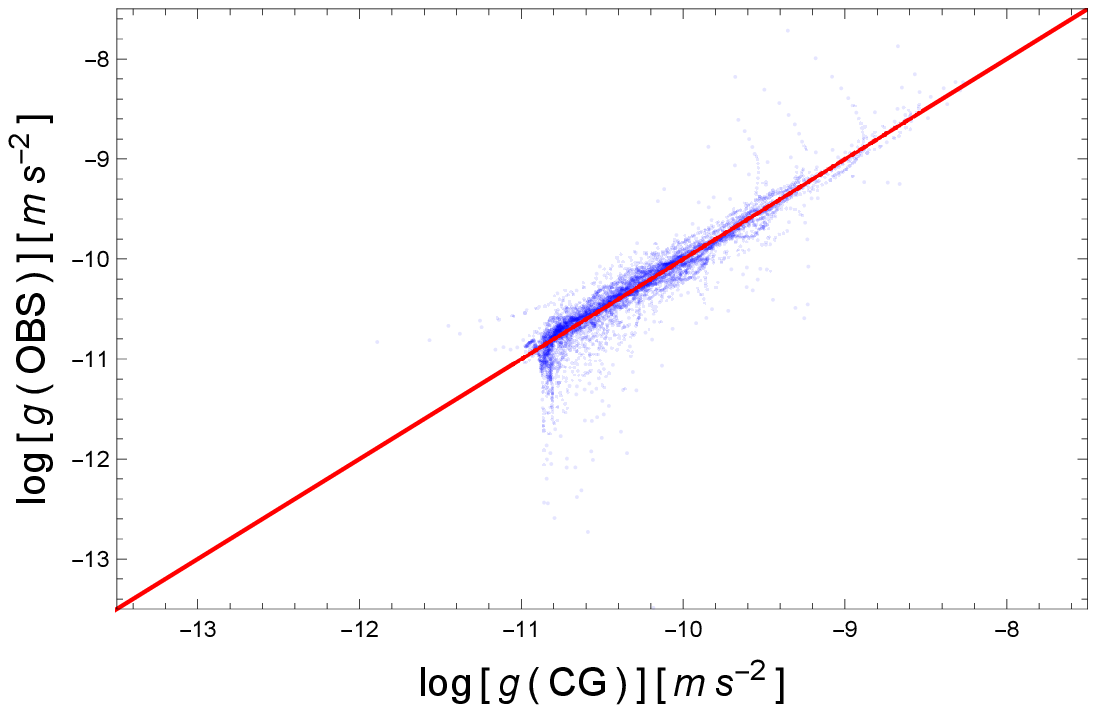,width=3.2in,height=2.05in}
		\label{opacity}}
	\subfigure[]{%
		\epsfig{file=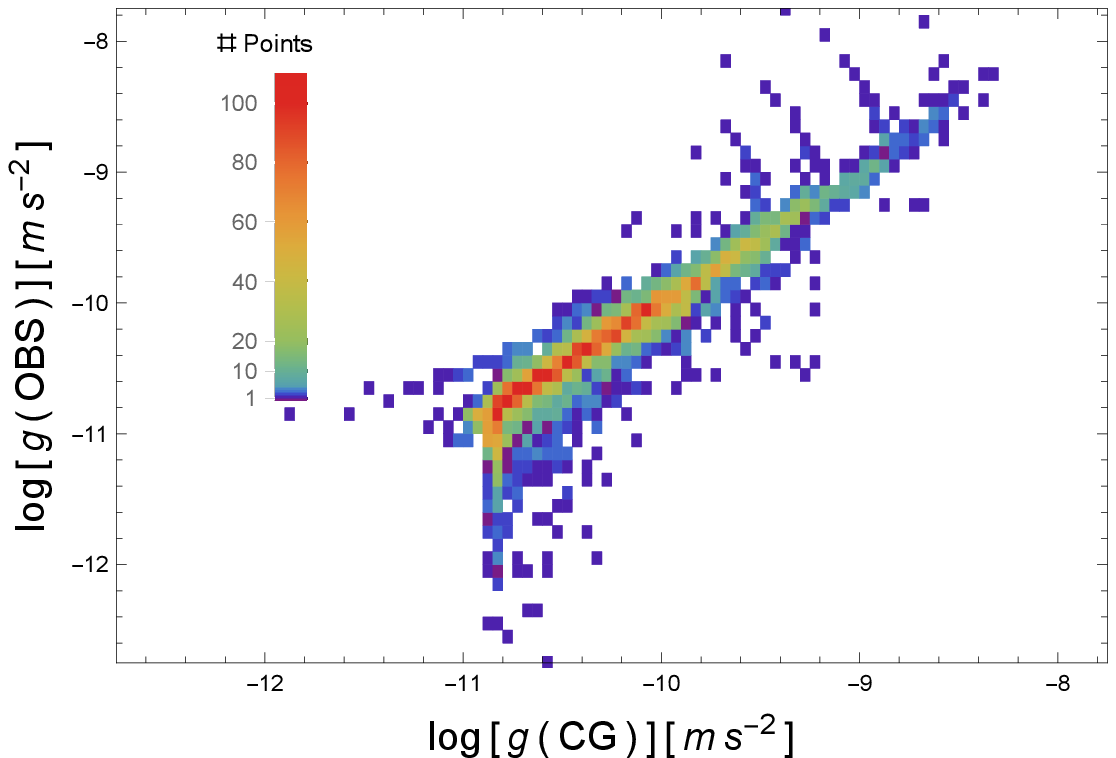,width=3.2in,height=2.05in}
		\label{density}}
	\caption{ Fig. \ref{opacity} shows the $(g_{OBS},g_{CG})$ plot, with the solid line being $g_{OBS}=g_{CG}$. Fig. \ref{density} shows a density binning of the points in Fig. \ref{opacity}.}
	
	\label{cgplots}
\end{figure}

Fig. 1  illustrates the consistency of the original McGaugh et al. data sample with our  data sample.  Fig. \ref{complot} shows the shows the exact data set as taken from the SPARC database while Fig. \ref{ourplot} shows all the data in our sample.  In  Figs. 1(a) and 1(b) we display eq. (\ref{E1}), to show  the correlation between the observed centripetal accelerations and the luminous Newtonian expectations. In the standard paradigm it is thought that dark matter is responsible for the fact that $g_{OBS}$ is not equal to $ g_{NEW}$ for low $g_{NEW}$.  In CG however, rotation curves are fit without dark matter, with the extra terms in eq. (\ref{total}) capturing the needed alternate physics.  Fig. \ref{cplot} shows the CG prediction on a point by point basis overlaid on the data in Fig. \ref{ourplot}. It can be seen from this plot that the CG fits fill out the entire width of the plot without the need for a fitting function such as eq. (\ref{E1}) that would approximate the data by a single curve.  To show that CG does fit the data, in Fig. \ref{cgplots} we plot $g_{OBS}$ against $g_{CG}=v^2_{CG}/R$.  We see that when CG is used for the fitting we obtain a one to one correlation across the entire data set without any adjustment of free parameters. \\

Fig. \ref{opacity}  shows that there are some points which are off the line of unity. 
The vertical descending lines in Fig. \ref{opacity} are due to some low rotational velocities at large distances in some dwarf galaxies (see for example \cite{lthingsapj}). There are also some vertical ascending lines in Fig. \ref{opacity} that represent scatter in the data.  To illustrate the negligible significance of the points that are off the line $g_{OBS}=g_{CG}$ in  Fig. \ref{density} we show a heat map of Fig. \ref{opacity} with cooler colors representing low counts of points, and hotter colors representing high counts.  This convenient visualization shows the dominance of the points on or near the line $g_{OBS}=g_{CG}$, and allows for binning of the data points.  Out of the total 6377 points, 205 data points fall into the descending or ascending categories described above, which amounts to a total of $3.2\%$ of the total sample.  Hence, conformal gravity can predict the observed centripetal accelerations to a high degree of accuracy in $97\%$ of the studied rotation curve data points.  
Fig. \ref{cgplots}  uses no additional free parameters to create these plots, and uses only the physics embodied in eq. (\ref{total}).  The CG theory provides agreement with the Fig. \ref{cgplots}  line of unity $g_{OBS}=g_{CG}$, with no dark matter being required at all.

\section{Tully-Fisher Relation}

The Tully-Fisher (TF) relation is an empirical relationship between the total luminous masses ($M=M_{disk}+M_{gas}+M_{bulge}$) and the observed rotation velocities as inferred from HI line widths.
With the quadratic term in eq. (\ref{total}) being negligible except at the largest radii, on evaluating the velocity at the point where the Newtonian Kepler fall off crosses the CG linear potential rise, 
we obtain a formula for the predicted $v^4$ at the crossover point of the form
 \begin{equation}
 v^4=\frac{BM}{M_{\odot}}\left(1+\frac{N^*}{D}\right),
 \label{cgtf}
 \end{equation}
where $B=2c^2M_{\odot}G\gamma_0=0.0074$ km$^4$s$^{-4}$ , $D=\gamma_0/\gamma^*=5.65\times10^{ 10}$, and $N^*=M/M_{\odot}$ is the number of stars in the galaxy in solar mass units.   As we see, the conformal gravity theory fixes the coefficient $B/M_{\odot}=2c^2G\gamma_0$ entirely in terms of fundamental quantities.

\begin{figure}[H]
  \centering
 \epsfig{file=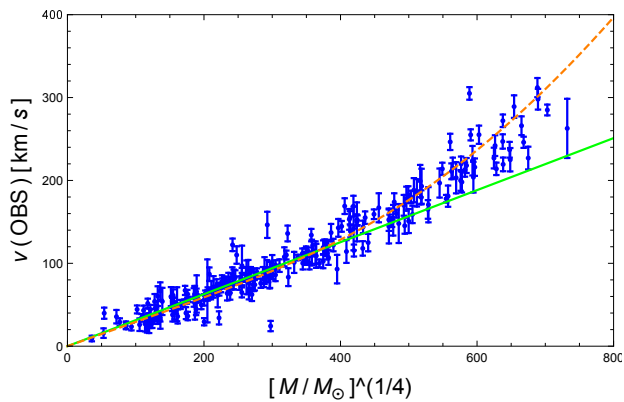,width=3.2in,height=2.05in}
  \caption{The Tully-Fisher plot for the 236 last data points of our sample as evaluated using the CG fitted masses $M$. Points are displayed with their reported velocity error. The solid line is the curve $v_{OBS}/M^{1/4}={\rm constant}$, while the dotted line shows the CG prediction of eq. (\ref{cgtf}).}
   \label{tullycg}
\end{figure}

For small dwarf galaxies $N^*$ is small compared to $D$, and for larger galaxies $N^*$ is of the order of $D$, leading to $BM/M_{\odot}<v^4<2BM/M_{\odot}$. With there being little difference between the velocities at the crossover points and the velocities at the last data points, eq. (\ref{cgtf}) then shows that the TF relation is a natural consequence of conformal gravity.  In Fig. \ref{tullycg} we construct the Tully-Fisher plot for the convenient last data point for all 236 galaxies in our data sample using the observed velocities and the fitted CG masses.  
In the figure we have plotted $v_{OBS}$ against  $M^{1/4}$ and incorporated the reported errors in the observed rotational velocities.  We overlay both $v/M^{1/4}={\rm{constant}}$   and eq. (\ref{cgtf}).
Fig. \ref{tullycg} shows that for most galaxies the CG eq. (\ref{cgtf}) is obeyed.  We note a potentially testable departure from the pure $v/M^{1/4}={\rm{constant}}$ and the CG eq. (\ref{cgtf}) at the largest galactic masses.  As we have noted above, accurate measurements of distance are imperative in rotation curve physics.  If the distance is incorrect, we would have a scaling issue and points on Fig. \ref{tullycg} would shift.  Moreover, in many galaxies, the TF relation is used to determine the distance in the first place.  This leads to a sometimes circular argument where the distance derived using the TF relation is used to make mass models which are then shown to obey the TF relation.  In future work, particular galaxies for which the TF relation is used to determine distance will be compared to other galaxies that use alternative distance measurements such as cepheids, in order to isolate the overall effects on modeling.

\section{Conclusion}
In McGaugh et al. \cite{mcgaughprl} the RAR  in galactic rotation curves was established via a set of 2693 total points. In this work we have shown that conformal gravity can universally fit the $g_{OBS}$ versus $g_{NEW}$ data in an even larger 6377 data point sample.  Conformal gravity has successfully fitted over 97$\%$  of the 6377 data points across 236 galaxies without any filtering of points, fixing of mass to light ratios or modification of input parameters.  Further, conformal gravity is shown to satisfy the $v^4\propto M$ relation consistently found in rotation curve studies, while also providing a derivation and extension of the TF relation.  We conclude by noting that there is a great deal of universality in rotation curve data. This universality does not obviously point in favor of dark matter, and is fully accounted for by the alternate conformal gravity theory.  

\ack
J.G. O'Brien would like to thank the International Association of Relativistic Dynamics (IARD) for the opportunity to present this work at the IARD 2018 Conference in Merida, Mexico.  J. G. O'Brien would like to thank J. Gagnon and Dr. P. Hafford for their help and support.

\bibliographystyle{iopart-num}
\bibliography{citationsnew}

\end{document}